\documentclass[%
 reprint,
 amsmath,amssymb,
 aps,
]{revtex4-1}

\usepackage{graphicx}% Include figure files
\usepackage{dcolumn}% Align table columns on decimal point
\usepackage{bm}% bold math
\usepackage{hyperref}
\hypersetup{
    colorlinks=true,
    linkcolor=blue,
    filecolor=blue,      
    urlcolor=black,
    pdftitle={Overleaf Example},
    pdfpagemode=FullScreen,
    }% add hypertext capabilities
\begin{document}

\preprint{APS/123-QED}

\title{Meta-lenses for differential imaging based on weak measurement}% Force line breaks with \\

\author{Xiong Liu}
\author{Rongchun Ge}
\email{rcge@scu.edu.cn}
\author{Xinrui Li}
\author{Jinglei Du}
\author{Hong Zhang}%
\author{Zhiyou Zhang}%
\email{zhangzhiyou@scu.edu.cn}
\affiliation{%
 College of Physics, Sichuan University, Chengdu China, 610064}%

\date{\today}% It is always \today, today,

\begin{abstract}
All-optical information communication, processing and computation have received substantial interest of both fundamental and applied research due to its unrivaled speed and broad bandwidth. Compared to its electronic counterpart, photons seldom interact with each other which makes them obtain a long coherence time on one hand and relieved from heavy energy dissipation on the other. However, one of the hindrances to achieve all-optical circuits is the large volume of all-optical devices to achieve specific functionalities. In this work, we propose and demonstrate experimentally three meta-lenses for differential imaging employing the framework of weak measurement: (1) partial differential lens, (2) total differential lens and (3) second order differential lens compatible with the requirement of miniaturization to achieve all-optical technology. Based on Fresnel-lens-like structures, our meta-lenses incorporated the previous weak-measurement compartment into wavelength scale, which induces a miniature differential operation system as a result. In addition to its potential importance in heavily integrated all-optical neural networks, the differential lens can be easily incorporated in the existing imaging systems like a conventional lens without increasing the complexity of the system of interest.
\end{abstract}

\maketitle

%\tableofcontents

\section{Introduction}

Optical communication has deeply transformed our world and our daily life due to its high speed, broad bandwidth, and extremely low energy dissipation in the past few decades \cite{kaushal2016optical,khalighi2014survey,li2009recent,Hanson:08}. Since then, there has been a tremendous effect to achieve all-optical information processing and computation in both fundamental research field and the research and development sector. The indispensable elements to achieve all-optical technology are all-optical devices realizing specific but basic functionalities such as differential operation.

Differentiation is a fundamental gate of operation in electronic circuits. It can be employed to obtain the differential image which contains the important boundary information of objects \cite{canny1986computational,marr1980theory} and forms the basis of applications, e.g., target detection, feature classification and data compression. In addition, in the artificial intelligence (AI) application of multilayer neural network, differential operation is widely implemented to realize algorithm optimization.  However, as the accessed data increases explosively just as in the rapid developing field of AI, it becomes increasingly expensive in terms of both time- and energy-consuming \cite{lecun2015deep}. In  contrast, photons due to their sharp wave-like property and the inertia to travel fast while stay free enable high-speed and energy-efficient computation for optical parallel analog computing free from the painstaking analog-to-digital and digital-to-analog convertors \cite{zangeneh2021analogue,abdollahramezani2020meta}. But a potential precariousness is that the original image received by the detector may be distorted by oversaturation  as an intensive beam is collected. Then, a preloaded differential imaging device that can strongly suppress the intensity of the incident light beam right before the detector provides a solution, which requires directly optical differential operation rather than electronic computation. It has been shown that, as a real-time operation, optical differentiation brings new alternative opportunities for fields like medicine and satellites \cite{pham2000current,holyer1989edge} in addition to the conventional electronic methods. Recently, physical imaging based on optical analog differential operation has drawn considerable attention \cite{silva2014performing,hwang2016optical,zhu2017plasmonic,fang2017grating,doskolovich2014spatial,youssefi2016analog,pors2015analog,golovastikov2015spatial,ruan2015spatial,guo2018photonic,fu2022ultracompact,wang2022single} due to its unique property complementary to its electronic counterpart; among which, the metamaterial scheme shows the promise of miniaturizing the size of the operation system \cite{silva2014performing,pors2015analog,hwang2016optical,fu2022ultracompact,wang2022single} -- a necessary step to achieve all-optical technology. And the relevant technique has been successfully applied to spin Hall effect of light and geometric phase to spatial differential \cite{zhu2019generalized,zhu2020optical,xu2020optical,he2020spatial,zhou2021two,zhou2019optical,xu2021enhanced}.

Weak measurement \cite{aharonov1988result} is an important research tool to tackle the statistical properties of a quantum system with pre- and post-selections measurement. It has been widely applied in exciting fields, such as precision estimations \cite{brunner2010measuring,starling2010continuous,xu2013phase,qiu2017precision, hosten2008observation,dixon2009ultrasensitive,magana2014amplification,wang2020measurement,xia2020high,liu2021high,fang2019improving,fang2016ultra,chen2021beating,yang2020experimental,yin2021improving}, ultrasensitive sensors \cite{li2017molecular, luo2017precision,zhang2016optical,li2019high,wang2020experimental}, fundamental physics research \cite{ yokota2009direct,lundeen2009experimental,brunner2004direct, lundeen2011direct,pan2019direct,liu2020experimental} and so on.  Recently, we have proposed to achieve differential imaging employing weak measurement, and experimentally demonstrated an effective partial differential imaging system \cite{liu2022general}. In this general scheme, a pre-selection initializes the system by selecting the input state, and this serves as the operating handle of the system, to which input function can be loaded. The weak coupling interaction introduces path-dependent evolution due to the energy splitting among these paths. Finally, in order to achieve the differential operation a proper post-selection is chosen to produce the desired geometric phase differences among the eigenstates. Unfortunately, like other strategies employed in current experiments, the system involved is usually large in volume.

\begin{figure}[htbp]
\centering
\includegraphics[width=8.5cm] {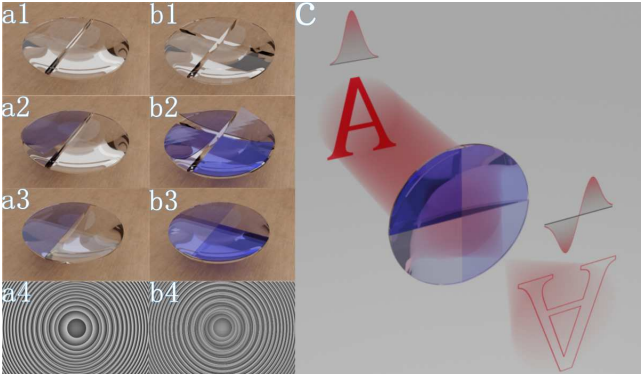}
\caption{\label{Fig1} Differential lenses (DLs) and differential imaging by DLs. Basic designing idea of partial differential lens (PDL) (a1-a3): a lens is cut into two identical semi-lenses; then a phase difference of $\pi$ and a small split ($\delta$) between them are introduced; finally both semi-lenses are glued together. For total differential lens (TDL) (b1-b3): a lens is cut into four identical quarter-lenses, and a distribution phases $0,\pi/2,\pi,3\pi/2 $ are imposed onto these quarter-lenses; a small split ($\delta$) between each adjacent two quarter-lenses is introduced. Experimental setups of PDL (a4) and TDL (b4): the specific PDL and TDL are realized in experiment by loading meta-lenses (a4) and (b4) onto a spatial light modulator, respectively. The Schematic of DLs imaging (c): Our DLs acting as a spatial differentiator that transforms an image into its partial and total derivative.}%Partial differential lens (PDL) and total differential lens (TDL). Basic designing idea of PDL (a-c): a lens is cut into two identical semi-lenses; then a phase difference of $\pi$ and a small split ($\delta$) between them are introduced; finally both semi-lenses are glued together. For TDL (d-f): a lens is cut into four identical quarter-lenses, and the the phases of $0,\pi/2,\pi,3\pi/2 $ are introduced in turn into these quarter-lenses; then a small split ($\delta$) between each adjacent two quarter-lenses is introduced; finally all the quarter-lenses are glued together. Experimental setups of PDL (g) and TDL (h): the specific PDL and TDL are realized in experiment by loading (g) and (h) onto a spatial light modulator, respectively.

\end{figure}

In this work, we introduce meta-lens to achieve the miniaturization of differential imaging system. Employing the generic weak measurement protocol \cite{liu2022general}, two kinds of differential lenses (DLs) are designed and demonstrated experimentally: (1) partial differential lens (PDL) and (2) total differential lens (TDL). The basic idea of our design is demonstrated in Figs.\ref{Fig1} (a1-a3) and Figs.\ref{Fig1}(b1-b3), where a relative phase difference of $\pi$ or $\frac{\pi}{2}$ has been induced among the (two/four) pieces of components for the PDL and TDL respectively. In our experiment, the DLs are realized employing Fresnel-lens-like structures (shown in Figs.\ref{Fig1}(a4, b4)), and the phase difference required is obtained via space light modulator (SLM). As a result, the length scale of DLs can be brought down to wavelength order (about $632.8nm$ in our experiments); such a miniaturization will save room for high integrated chips to finally achieve all-optical technology. By imposing the DLs upon imaging platforms like a conventional lens, as shown in Fig.\ref{Fig1}(c), a differential image of the input signal is obtained. Consequently, efficient signal/information compressing is achieved when only the information of the edges is vital, which could be exploited to relieve the loading burden of high-sensitive detector, and make real-time imaging processing and the transmission and storage of data easier. Moreover, the noise received by the detector can be suppressed too due to the presence of weak measurement. To show that such a design strategy is versatile, we also demonstrate briefly an experimental realization of a second order DLs and propose to achieve a third order DLs.

\section{New design strategy of differential lenses}
 Our DLs are based on the generic weak measurement protocol. For PDL, the operating handle is a achieved by a two-level ancilla preselected as $|\Phi_{pre}\rangle=\dfrac{1}{\sqrt{2}}(|0\rangle+|1\rangle)$; here $|0\rangle$, $|1\rangle$ are the eigenstates of its observable $\hat{A}=|0\rangle\langle0|-|1\rangle\langle1|$, and are physically corresponding to the two different evolution paths following the two semi-lenses (Figs.1(a-c)). The system of interest is the incident wave of the input image ($\phi(x,y)$), which is weakly coupled to the ancilla. The weak interaction can be described as 
$\hat{U}=\exp(-\delta\hat{A}\dfrac{\partial}{\partial x})$, which will introduce a small splitting $2\delta$ between the two paths $|0\rangle$ and $|1\rangle$. After passing the PDL, the state of the whole system becomes
\begin{align}
|\Psi(x,y)\rangle&=\hat{U}|\Phi_{pre}\rangle\phi(x,y)\nonumber\\
&=\dfrac{1}{\sqrt{2}}\Big(\phi(x-\delta,y)|0\rangle+\phi(x+\delta,y)|1\rangle\Big)\label{e1}. 
\end{align}
By our designing a relative phase different of $\pi$ is induced between the semi-lenses, and this will achieve a post-selection as
\begin{align}
|\Phi_{post}\rangle&=\dfrac{1}{\sqrt{2}}(|0\rangle-|1\rangle)\label{e2}. 
\end{align}
Such a selection of phase difference is employed to make sure that the post-selected state of the ancilla is orthogonal to its initial state; a generalization of the present design can be achieved without additional complexity as will be shown bellow for the TDL. As a result of the post-selection, the system of interest is obtained as
\begin{align}
    \phi_f(x,y)&=\langle\Phi_{post}|\Psi(x,y)\rangle\nonumber\\
    &=\dfrac{1}{2}\Big(\phi(x-\delta,y)-\phi(x+\delta,y)\Big).\label{e3}
\end{align}
As is general in weak measurement, the typical length scale of the initial wave function profile is much larger than the lens split ($\delta$), so the final output wave function is  well approximated by its first derivative of the input wave function of the image:
\begin{equation}
    \phi_f(x,y)=-\delta\dfrac{\partial}{\partial x}\phi(x,y).\label{e4}
\end{equation}
Therefore,  in the Fourier transformed $k-$space the output intensity received by the detector is  $(\delta k)^2$ times the differential intensity of the input image. Compared with the input intensity, the output intensity is greatly compressed as  $(\delta k)^2\ll1$, which prevents the detector from being oversaturated by too strong light. 

For TDL, the preselected state is $|\Phi_{pre}\rangle=\dfrac{1}{2}\sum_{mn}|x_m,y_n\rangle$, where $|x_m,y_n\rangle$ with $m,n=\pm1$ being the four eigenstates corresponding to the four paths of TDL (Figs.1(d-f)). The  weak interaction between the four-level ancilla and the input function $\phi(x,y)$ can be described by  $\hat{U}=\exp(-\delta\hat{A}(\dfrac{\partial}{\partial x}+\dfrac{\partial}{\partial y}))$,  with observable of the ancilla $\hat{A}=|x\rangle\langle x|+|y\rangle\langle y|$ where $\hat{A}|r_\pm\rangle=\pm|r_\pm\rangle$ ($r$ is the coordinate $x$ or $y$). The whole system after the weak interaction is
\begin{align}
|\Psi(x,y)\rangle=&\hat{U}|\Phi_{pre}\rangle\phi(x,y)\nonumber\\
=&\dfrac{1}{2}\sum_{m,n=\pm}\phi(x-m\delta,y-n\delta)|x_m,y_n\rangle\label{e5}. 
\end{align}
Equation (\ref{e5}) shows that the wave function in four paths of TDL obtains a tiny spatial shift ($\pm\delta$) depends on the directions. Here the relative phase differences between adjacent quarter- lenses are $\pi/2$. The output wave function after the post-selection is 
\begin{align}
   \phi_{f}(x,y)=&-\dfrac{1}{2}\delta[\dfrac{\partial}{\partial x}\phi(x,y)+\dfrac{\partial}{\partial y}\phi(x,y)]\nonumber\\
   &+\dfrac{i}{2}\delta[\dfrac{\partial}{\partial x}\phi(x,y)-\dfrac{\partial}{\partial y}\phi(x,y)],\label{e6}
\end{align}
and the intensity profile is
\begin{align}
   I_f(x,y)&=\phi^*_f(x,y)\phi_f(x,y)\nonumber\\
   &=\dfrac{1}{2}\delta^2[(\dfrac{\partial}{\partial x}\phi(x,y))^2+(\dfrac{\partial}{\partial y}\phi(x,y))^2].\label{e7}
\end{align}
 It is shown clear in Eq.~(\ref{e7}) the output intensity by TDL is differentiated in both directions which is just the total differential image of the input profile. As is in the case of PDL, information compression is achieved with the same efficiency. As a generalization of the traditional 2-path weak measurement protocol, our 4-path design of the TDL provides a new way to design optical analog operating systems/circuits. Although our analysis presented above is for single photon, the same procedure is applicable to macroscopic beams which will be the case of our experiment following \cite{liu2022general}. 

\section{Experimental realization}

\begin{figure}[htbp]
\centering\includegraphics[width=8.5cm] {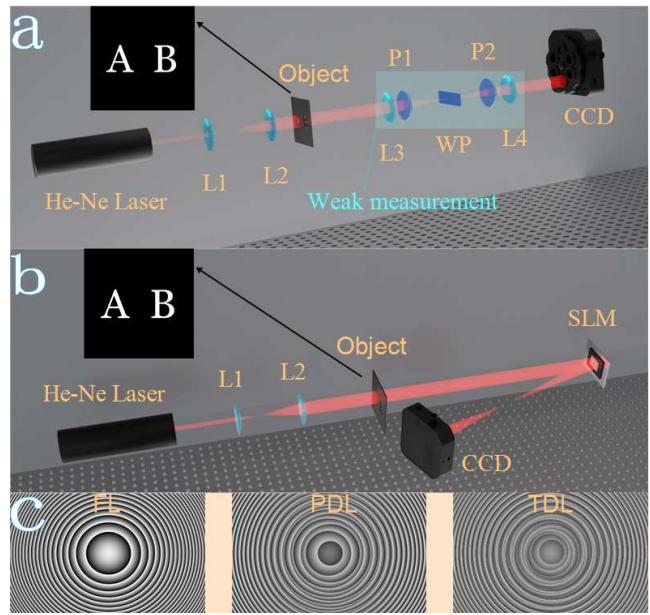}\caption{\label{Fig2} (a) Experimental setup of Wollaston prism based differential imaging platform, 
 where the weak measurement consists of five cascaded subblocks: (i) a Fourier transform lens (L3), (ii) a pre-selection (P1), (iii) a weak coupling (WP), (iv) a post-selection (P2), and (v) an inverse Fourier transform lens (L4). (b)Experimental setup of our new differential lenses imaging platform, where the SLM plays all the five roles of weak measurement. (c) Fresnel lens (FL) and our two meta-lenses (PDL and TDL). The light source is He-Ne laser (wave length $\lambda_0 = 632.8\ {\rm nm}$). L1 and L2, beam expander, P1 and P2, polarizers; Object, the input image; L3 and L4, lenses with focal lengths 250 mm; WP, an ultrathin Wollaston prism; SLM, spatial light modulator; CCD, Thorlabs BC106N-VIS.}
%Experimental setup: the light source is He-Ne laser (wave length $\lambda= 632.8\ {\rm nm}$); L1 and L2, beam expander; Object, the input image; P, polarizer; SLM, spatial light modulator; CCD, Thorlabs BC106N-VIS. FL, PDL and TDL are Fresnel lens, Partial differential lens and total differential lens to be loaded onto SLM.
\end{figure}

In this section,  we present an experimental realization of the DLs proposed in previous section employing Fresnel lens (FL). Our experimental setup is shown schematically in Fig.\ref{Fig2}(b) where SLM is used to induce the phase difference required by the DLs. Light emitted by the He-Ne laser is expanded by L1 and L2,  then it is projected to the screen of the object obtaining a specific spatial amplitude distribution image (letter 'A' or 'B') which is encoded as $\phi(x,y)$. Finally, the differential image is received at CCD. For comparison, in our weak measurement differential imaging platform based on ultra-thin Wollaston prism\cite{liu2022general}, the weak measurement system consists of two lenses (L3 and L4), two polarizers (P1 and P2) and an ultra-thin Wollaston prism, as shown in Fig.\ref{Fig2}(a); in the DL imaging system, the DL generated by the spatial light modulator has effectively compressed the five weak measurement processes into a single piece as shown in Fig.\ref{Fig2}(b).
 
\begin{figure}[htbp]
\centering\includegraphics[width=8.5cm] {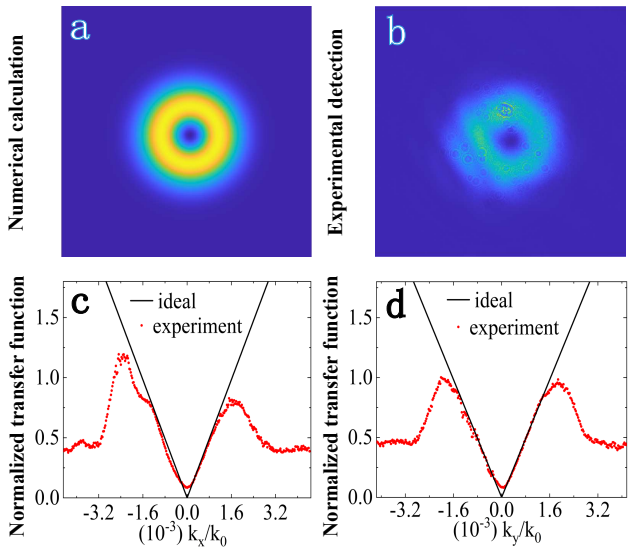}\caption{\label{Fig3} The spatial spectral transfer function and differential imaging of TDL with an input Gaussian beam. (a,b) are the results of the total derivative of the Gaussian beam obtained with simulation and experiment, respectively. (c,d). Spatial spectral transfer function of TDL for $k_y=0$ and $k_x=0$. Experimental results are shown in red dots, and theoretical results are in black lines.}
\end{figure}

 For reference, an image with Gaussian distribution whose differential image is well known analytically is tested for both our PDL and TDL. In order to eliminate the influence of the  Fresnel lens being loaded onto the SLM, the output of the FL is taken as our standard input image for comparison. The experimental data of FL and PDL are consistent with the results of our weak measurement differential imaging platform based on an ultra-thin Wollaston prism \cite{liu2022general}. And when TDL is loaded on the SLM, a circular beam pattern is received at CCD, as shown in Figs.~\ref{Fig3}(b). For comparison to theory, the output of the TDL is numerically fitted with the total differential image of a Gaussian profile, as shown in Fig.~\ref{Fig3}(a), it is a circular beam with center symmetry, while the experimental beam has a small fringe on the left side, which is the interference streak caused by the lack of the phase accuracy of SLM.

In order to compare the results of experiment to numerical simulation, the differential transfer functions are calculated for both DLs. For the output of the CCD, only the intensity of the field profile is recorded, as a result the transfer functions of the DLs employed can be formulated as \cite{liu2022general}
\begin{equation}
    H(k_x,k_y)=\dfrac{\sqrt{I_o(k_x,k_y)}}{\sqrt{I_i(k_x,k_y)}}=\dfrac{\sqrt{I_o(x,y)}}{\sqrt{I_i(x,y)}},\label{e8}
\end{equation}
where $I_o$ and $I_i$ are the output (differential images) and input (original image) intensities, respectively. Our experimental data of the transfer function for PDL agrees well with our previous result using a different imaging setup \cite{liu2022general} and with numerical simulation too. For the TDL, the transfer function yields a 2-dimensional distribution as shown in Fig.~\ref{Fig3}(c, d): our experimental data and numerical simulation show excellent agreement with each other in the central region; but a large derivation is observed for large momentum which physically corresponds to the tails of the signal where the intensity is weak. Since the $x$ or $y$ axes selected above are of no special, it can be inferred that TDL works in any direction. We notice that there exists an unexpected small stripe in the left region of Fig.\ref{Fig3}(b), which led to an exception in the transfer function, as shown in Fig.\ref{Fig3}(c), the left region shows a large peak relative to the right region. Due to the symmetry of the TDL in left and right regions, we think that the stripe can be avoided if a SLM of higher resolution is employed i.e., if the TDL distribution is smoother. Nevertheless, in the main central region of Gaussian light, the experimental transfer function agrees well with the standard numerical calculations.

\begin{figure}[htbp]
\centering\includegraphics[width=8.5cm] {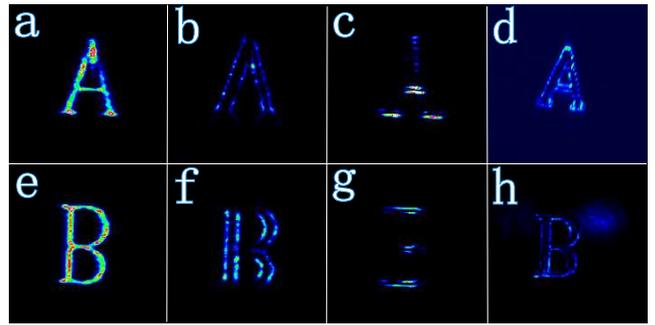}\caption{\label{Fig4} Images and the first order differential images of letters 'A' and 'B'. (a) and (e), input images with amplitude distribution of the letters 'A' and 'B', respectively. (b) and (f), the first order derivatives of letters in x-direction by PDL. (c) and (g), the first order derivatives of letters in y-direction by PDL. (d) and (h), the first order derivatives of letters in two-dimensions by TDL.}
\end{figure}

To show that our design of the DLs works in general for non-trivial geometry configurations, the objects with amplitude distributions of letters A and B are tested, of which the outline is uniformly distributed, while the rest is dark. However, the images of letters A and B produced by FL loaded on the SLM are slightly distorted, as shown in Figs.\ref{Fig4}(a) and \ref{Fig4}(e). We attribute this to Fresnel lens and the large pixel size of SLM (about $64\mu m^2$), which will have an impact on the profiles of the correspondent differential images. Nevertheless, the images obtained in experiment are sufficient to demonstrate the effect of differential imaging of the DLs. As shown in Figs.\ref{Fig4}(b) and \ref{Fig4}(f), the differential operation performed by PDL is along x-direction, so the edges perpendicular to x-direction are most visible while the x-direction are invisible. Furthermore the edge of input beam can be detected as long as it is not completely along x-direction. The similar thing happens in the y-direction as shown in Figs.\ref{Fig4}(c) and \ref{Fig4}(g), the edges perpendicular to y-direction are most visible while invisible in y-direction. Consequently, by rotating the PDL, the differential operations can be performed in any direction of interest. As for our TDL, the total differential images of letters A and B obtained are shown in Figs.\ref{Fig4}(d) and \ref{Fig4}(h); the edges of both directions are shown clearly, which indicate that the TDL successfully yields all the edge information of the original images. In other words, we achieve optical total differential operation of the object through a wavelength thick TDL.

\section{Discussion and Conclusion}

\begin{figure}[htbp]
\centering\includegraphics[width=8.5cm] {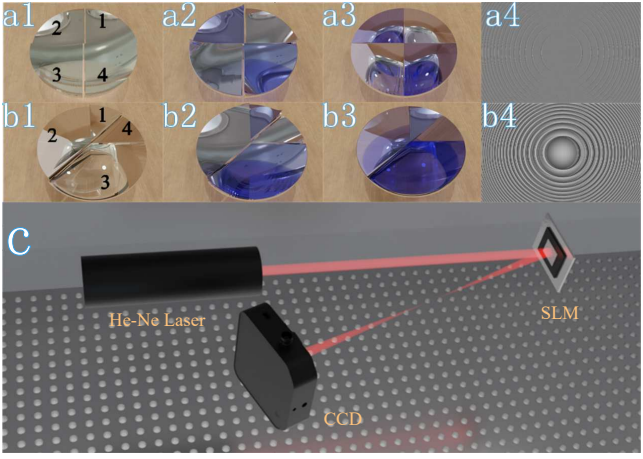}\caption{\label{Fig5} Higher order DLs and the experimental setup of the second order differential imaging. The second order partial differential lens (SOPDL) (a1-a3): a lens is cut into four identical quarter-lenses labeled with the numbers 1, 2, 3, 4 (a1), and the relative phases of $0, \pi, 0, \pi$ are exposed onto these quarter-lenses; then the 1st and 3rd pieces are shifted along $x$-direction by $\delta$ and $-\delta$, respectively. At the end, all pieces are glued together. The third order partial differential lens (TOPDL) (b1-b3): a lens is cut into four unequal lenses labeled with the numbers 1, 2, 3, 4 (b1) with phases of $0, \pi, 0, \pi$ introduced; these lenses are then shifted along $x$-direction, by $\delta$, 0, $-\delta$, $-2\delta$ in turn. Experimental setups of meta-SOPDL (a4) and the scheme for meta-TOPDL (b4) by loading a spatial light modulator. 
Experimental setup of the SOPDL imaging platform (c) with a Gaussian beam generated by He-Ne laser; the SOPDL is represented by a SLM screen loaded with (a4); and CCD receive the outgoing image.}
\end{figure}

Our differential lenses based on weak measurement have successfully miniaturized an all-optical differential imaging system down to wavelength scale. Aside from this, our two- and four-path design for the DLs is ready to be adapted into a multi-path design for new optical devices which will open a new door in the designing all-optical devices. As a proof of principle, we show briefly an experimental realization of a second-order partial differential lens (SOPDL) and a proposal for third-order partial differential lens (TOPDL), as shown in Figs.\ref{Fig5}(a1-a3) and Figs.\ref{Fig5}(b1-b3). For SOPDL, the pre-selected state is the same of TDL, $|\Phi_{pre}\rangle=\dfrac{1}{2}\sum_{mn}|x_m,y_n\rangle$; where $|x_m,y_n\rangle$ with $m,n=\pm1$ being the four eigenstates corresponding to the four paths of SOPDL (Figs.\ref{Fig5}(a1-a3)). The  weak interaction is introduced in paths $|x_+,y_+\rangle$ and $|x_-,y_-\rangle$, so the input function $\phi(x,y)$ can be described by $\hat{U}=\exp(-\delta\hat{A}\dfrac{\partial}{\partial x})$,  with $\hat{A}=\sum_{m=\pm}|x_m,y_m\rangle\langle x_m,y_m|$ observable of the ancilla; where $\hat{A}|x_m,y_m\rangle=\pm|x_m,y_m\rangle$ and $m=\pm1$. The whole system after the weak interaction is
\begin{align}
&|\Psi(x,y)\rangle=\hat{U}|\Phi_{pre}\rangle\phi(x,y)\nonumber\\
=&\dfrac{1}{2}\sum_{m=\pm}[\phi(x-m\delta,y)|x_m,y_m\rangle+\phi(x,y)|x_m,y_{-m}\rangle]\label{e9}. 
\end{align}
 Eq.~ (\ref{e9}) shows that the wave function in the paths $|x_m,y_m\rangle$ ($m=\pm1$) of SOPDL obtain a tiny spatial shift $\pm\delta$, while the wave function in the rest paths ($|x_m,y_{-m}\rangle$ with $m=\pm1$) remains the same. Here the relative phase difference between adjacent quarter lenses is $\pi$. The output wave function after the post-selection is 
\begin{align}
   \phi_{f}(x,y)=&\dfrac{1}{2}[\phi(x+\delta,y)-\phi(x,y)+\phi(x-\delta,y)-\phi(x,y)]\nonumber\\
%   =&\dfrac{1}{4}[\delta\dfrac{\partial}{\partial x}\phi(x+\dfrac{1}{2}\delta,y)-\delta\dfrac{\partial}{\partial x}\phi(x-\dfrac{1}{4}\delta,y)]\nonumber\\
   \approx&\dfrac{1}{4}\delta^2\dfrac{\partial^2}{\partial x^2}\phi(x,y),\label{e10}
\end{align}
with the profile of intensity given by
\begin{align}
   I_f(x,y)%&=\phi^*_f(x,y)\phi_f(x,y)\nonumber\\
   &=\dfrac{1}{16}\delta^4(\dfrac{\partial^2}{\partial x^2}\phi(x,y))^2.\label{e11}
\end{align}
 So the output field of SOPDL is the second-order differentiation of the input profile.

\begin{figure}[htbp]
\centering\includegraphics[width=8.5cm] {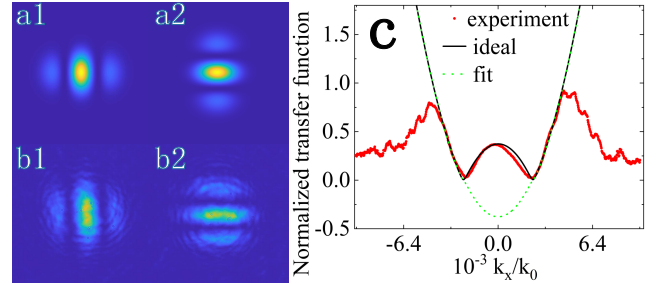}\caption{\label{Fig6} The second order derivative images in $x$, and $y$, axes for an input Gaussian beam. (a1-a2), the simulated distributions. (b1-b2), the experimental distributions. (c), the transfer function of SOPDL in $x$ direction. %(h) Spatial spectral transfer functions of SOPDL for $k_y=0$, 
The green dotted line is a quadratic fit. Experimental results are shown in red dots, and theoretical results are in black lines.}
\end{figure}

A Gaussian beam generated by the laser has been employed as a proof of principle for our SOPDL. Fig.~\ref{Fig6}(b1) shows the differential image of the input Gaussian beam in $x$-direction literally as in Eq.~(\ref{e11}). As for comparison, the simulated results with an ideal Gaussian shape is shown in Fig.~\ref{Fig6}(a1). By rotating the SOPDL, the second order differential image of Gaussian input in $y$-direction can be obtained easily as shown in Fig.~\ref{Fig6}(b2) with the correspondent numerical simulation in Fig.~\ref{Fig6}(a2).  The transfer function in $x$ and $y$ directions of SOPDL are equally distributed except for the orientation, so we show only the situation of $x$ direction in Fig.~\ref{Fig6}(c). Experimental data and numerical simulation agree well with each other in the central region, but a large derivation is still observed for large momentum which physically corresponds to the tails of the signal where the intensity is weak. The transfer function of SOPDL can be fitted with a quadratic function, $H(k_x)\propto k_x ^2$, as shown in Fig.~\ref{Fig6}(c), green dashed line. Since the field obtained from the light intensity is not negative, the quadratic function is opposite to the experimental data in the negative segment.

 As a last demonstrate of the versatility of our strategy, we propose the achieve TOPDL. Here a lens is cut into four unequal pieces: $|1\rangle$, $|2\rangle$, $|3\rangle$ and $|4\rangle$ as shown in (Figs.\ref{Fig5}(b1-b3)). The pre-selected state is  $|\Phi_{pre}\rangle=\dfrac{\sqrt{2}}{4}|1\rangle+\dfrac{\sqrt{6}}{4}|2\rangle+\dfrac{\sqrt{6}}{4}|3\rangle+\dfrac{\sqrt{2}}{4}|4\rangle$. The  weak interaction introduces relative shifts of $\delta$, 0, $-\delta$, $-2\delta$ (in $x$-direction) in the four paths of TOPDL. In addition, a relative phase difference of $\pi$ is applied between adjacent lenses. The output wave function after the post-selection is
 \begin{align}
   \phi_{\rm f}(x,y)=&\sum_{i,j}^{0,1}\dfrac{1-(-1)^j2}{8}\phi(x-\frac{\delta}{2}-(-1)^{i+j}\frac{1+(-1)^i2}{2}\delta,y)\nonumber\\
   =&\delta\sum_{j=1}^3(-1)^j\frac{2-|j-2|}{8}\dfrac{\partial}{\partial x}\phi(x+\dfrac{3}{2}\delta-j\delta,y)\nonumber\\
   =&\delta^2[\dfrac{1}{8}\dfrac{\partial^2}{\partial x^2}\phi(x,y)-\dfrac{1}{8}\dfrac{\partial^2}{\partial x^2}\phi(x-\delta,y)]\nonumber\\
   =&\dfrac{1}{8}\delta^3\dfrac{\partial^3}{\partial x^3}\phi(x+\dfrac{1}{2}\delta,y),\label{e12}
\end{align}
and the correspondent profile of intensity is
\begin{align}
   I_{\rm f}(x,y)&=\phi^*_{\rm f}(x,y)\phi_{\rm f}(x,y)\nonumber\\
   &=\dfrac{1}{64}\delta^6(\dfrac{\partial^3}{\partial x^3}\phi(x+\dfrac{1}{2}\delta,y))^2.\label{e13}
\end{align}
 So the third-order differential image of an input profile can be obtained in principle with our TOPDL.

In summary, we have proposed a generic idea to achieving differential lens. Employing the basic framework of weak measurement, we presented an effective way of designing DLs and miniaturizing differentiating optical devices. We demonstrated experimentally the PDL, TDL, and SOPDL, and propose to achieve TOPDL %both the PDL and TDL 
based on Fresnel lenses. Our differential imaging lenses were experimentally achieved by loading meta-lenses onto the SLM. The differential imaging of Gaussian light was used to benchmark the transfer function of DLs. %By comparing with numerical simulation, our experimental results confirmed that 
The transfer functions of DLs were obtained experimentally and agrees well with numerical simulations.  As a non-trivial application, differential images of objects with amplitude distributions were shown experimentally for both PDL and TDL. The edge image of the object was obtained only in a specific direction with PDL, and in both directions with TDL. We demonstrated the differential images of objects with amplitude-distributed information only, but it works for objects with phase distribution too (for both the PDL and TDL) as these operations are carried over the field instead of the intensity. Moreover, the noise is suppressed and the data received by the CCD is compressed because of the partition of weak measurement which is integrated on the DLs.

In our specific realizations, the weak measurement compartment is integrated into the full imaging system for the first time by a single differential lens. This is accomplished by matching a finite phase depth of $0$-$2\pi$ of the SLM on top of Fresnel-lens-like structures. As a result, a miniaturization of the system on the wavelength scale is achieved. Because DLs are essentially lenses, they can be integrated into an imaging platform like a conventional lens, or simply replace an objective lens without incurring additional complexity to the  design of the imaging platform. Our two and four-path design based on weak measurement could be generalized to a generic multi-path scheme while keeping the merit of miniature,  as is further exemplified by the SOPDL and TOPDL. This will open a new door in the designing of all-optical device and contribute to the field of all-optical technology. %{\Red Moreover, the differential orientation of an object can be changed by simply rotating the differential lens.}
%From Eqs.(\ref{e4}), (\ref{e10}) and (\ref{e12}), we can obtain that the higher the differential order, the weaker the field, and light intensity is reduced by a multiple of $\delta^2$. As a result, when the SOPDL and TOPDL are loaded on the SLM, the outgoing light is too weak to be detected. We use the Gaussian beam generated by the laser as the input image without expanded in order to avoid unnecessary photon loss and obtain the second order derivative of Gaussian beam in $x$ direction. However, the outgoing light of TOPDL is still too weak to be detected, as shown in Eq.(\ref{e13}), the differential light intensity is reduced by a multiple of $\delta^6$.

In conclusion, we have reported an approach to realize all-optical differential operation using the framework of weak measurement. we believe that more compact computing lenses will appear in the future under the leads of weak measurement or not, and eventually be extended to artificial intelligence and high-performance computing of all-optical networks.

\begin{acknowledgments}
This work is supported by the National Natural Science Foundation of China (11674234), the Science Specialty Program of Sichuan University (2020SCUNL210), and by the Fundamental Research Funds for the Central Universities. 

\end{acknowledgments}

\section{appendix: how to design meta-differential lenses}
\begin{figure}[htbp]
\centering
\includegraphics[width=8.5cm] {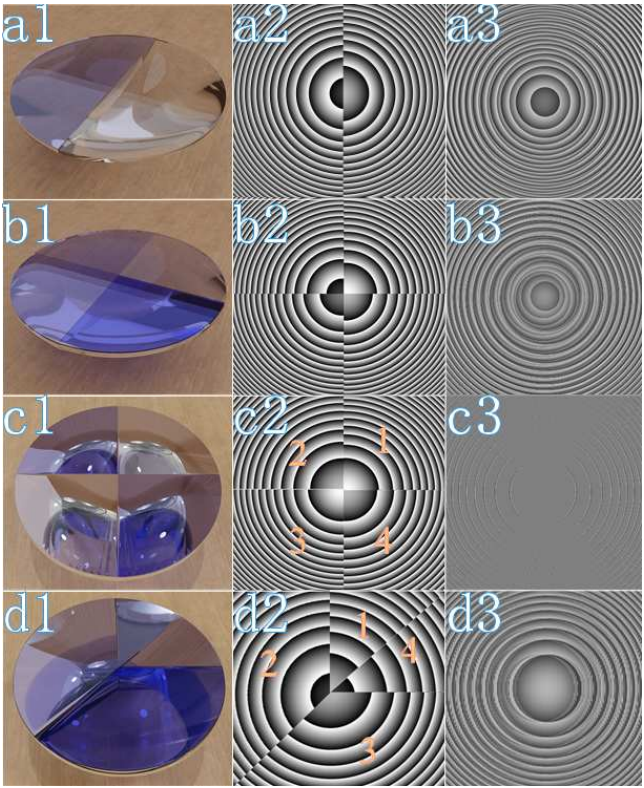}
\caption{The three generations of PDL (a1-a3), TDL (b1-b3), SOPDL (c1-c3) and TOPDL (d1-d3).}
\label{FigA}
\end{figure}

According to the idea of differential lenses, all the DLs %both the PDL and TDL 
are firstly designed as Figs.\ref{FigA}(a1-d1), respectively. However, in the experimental realizations, the SLM is used to represent the lenses. So the DLs should be designed like Fresnel-lens to be loaded on the SLM. In this idea, for PDL, firstly, we design two Fresnel-lenses with a phase difference of $\pi$ between them; secondly, one lens is cut off the right half while the other one the left half, i.e., they are cut into semi-Fresnel-lenses; finally, a small split ($\delta$) between them are introduced and both semi-lenses are glued together to be a PDL, As shown in Fig.\ref{FigA}(a2). It can also be understood that the left and right regions of the PDL are designed as distinct Fresnel-lenses with a phase difference of $\pi$ between them and a distance of $\delta$ between their centers, which represent two evolution paths of PDL. For TDL, there are four evolution paths, so four regions are designed as four different Fresnel-lenses with the phases differences of $0$, $\pi/2$, $\pi$, $3\pi/2$ and the centers of two adjacent lenses separated by the distance of $\delta$, as shown in Fig.\ref{FigA}(b2). Similar designs appear on SOPDL and TOPDL, but with different areas and phases, as shown in Fig.\ref{FigA}(c2, d2).

However, there is still a problem: each region of DLs represents one evolution path, according to the theory of DLs imaging, the light should equally pass through every evolution paths of DLs, what if the light passes through the paths unequally? Or even just passes through some regions instead of all the regions? To solve this problem, we further improve the design of DLs. We make every pixel of SLM represent a different evolution path. Specifically, for PDL, the odd column pixels on the SLM represent one evolution path, while the even column pixels represent another evolution path. In this way, the probability is greatly increased that the light travels equally through all the evolution paths at the same time. For TDL and SOPDL, the four adjacent pixels on the SLM respectively represent four different evolution paths. For TOPDL, the eight adjacent pixels respectively represent four different evolution paths, e.g., the first and fourth areas are represented by one pixel each, while the second and third areas are represented by three pixels each. Therefore, the final DLs are designed as meta-lenses as shown in Fig.\ref{FigA}(a3, b3, c3, d3). 

\nocite{*}

\end{document}